Hole mobility in organic single crystals measured by a "flip-crystal" field-effect technique

C. Goldmann, S. Haas, C. Krellner, K. P. Pernstich, D. J. Gundlach, and B. Batlogg

Laboratory for Solid State Physics

ETH Zurich, 8093 Zurich, Switzerland

Phone: + 41 1 633 2361, Fax: + 41 1 633 1072, Email: goldmann@solid.phys.ethz.ch

*Abstract* – We report on single crystal high mobility organic field-effect transistors (OFETs) prepared on prefabricated substrates using a "flip-crystal" approach. This method minimizes crystal handling and avoids direct processing of the crystal that may degrade the FET electrical characteristics. A chemical treatment process for the substrate ensures a reproducible device quality. With limited purification of the starting materials, hole mobilities of 10.7, 1.3, and 1.4 cm$^2$/Vs have been measured on rubrene, tetracene, and pentacene single crystals, respectively. Four-terminal measurements allow for the extraction of the "intrinsic" transistor channel resistance and the parasitic series contact resistances. The technique employed in this study shows potential as a general method for studying charge transport in field-accumulated carrier channels near the surface of organic single crystals.

*Introduction* – Charge transport in organic semiconductors has been a subject of fundamental study for decades.[1-3] In recent years interest in organic semiconductors for use as the active layer of thin film transistors has increased, and with improved material preparation and device processing the electrical performance of organic thin film transistors (OTFTs) is similar to that of hydrogenated amorphous silicon. The room temperature field-effect mobility of the best OTFTs now approaches what was thought to be theoretical limits for chemically and structurally perfect organic molecular crystals (OMCs). However, the temperature dependence of the current-voltage characteristics shows that even in the best OTFTs the intrinsic properties are masked by defects. Single crystals are therefore not only of scientific but also of technological interest, being ideally suited for investigating intrinsic charge transport properties and intrinsic device limitations.

Anthracene has served for decades as a prototype material for fundamental charge transport studies, and recently the homologues tetracene and pentacene have received considerable attention owing to their high mobility and their potential use in thin film organic electronics.[4-6] *Karl et. al.* have demonstrated that sample perfection (chemical and crystalline) and device preparation are of critical importance for accessing the intrinsic transport properties of organic molecular single crystals.[7] In recent work using tetracene single crystals *R. W. I. de Boer et. al.* report how the intrinsic transport properties can be unintentionally masked by defects introduced during sample processing.[8]

Historically time-of-flight (TOF) and space-charge-limited current (SCLC) measurements have been used for transport studies. While well suited for many inorganic semiconductors, limitations in the processibility of many OMCs often restrict these measurements to a single transport direction. For example, in pentacene and tetracene the transport direction most easily accessed with TOF and SCLC measurements corresponds to the c-axis. In this direction charge transport is poor due to very little pi-orbital overlap. In-plane transport studies of OMCs have proven to be more difficult using TOF and SCLC measurements with respect to both the sample fabrication and the interpretation of data.

Insulated-gate field-effect transistor (FET) structures are well-suited for in-plane studies since in these structures charge is accumulated and transported in a channel near the gate insulator/OMC interface. We report here on organic field-effect transistors (OFETs) fabricated from rubrene, tetracene and pentacene single crystals using a "flip-crystal" method which minimizes crystal handling. We first discuss four-terminal measurements on rubrene in detail, and then give an overview of the results obtained for all three materials mainly by two-terminal measurements. These results are comparable or better than those recently reported for the same materials in various FET-configurations.[8-13] They demonstrate the good reliability and reproducibility of the "flip-crystal" characterization method and its usefulness for investigating the properties of OMCs. Additionally, we address issues that remain problematic and are under further study.

*Experimental* – OFETs were fabricated using an advanced form of the structure and process developed and reported on previously.[10] This "flip-crystal" method minimizes the handling of the thin fragile organic crystals and reduces the likelihood of damaging their surface, which is always a concern when metal contacts or a dielectric layer are deposited on the crystal surface. Heavily doped thermally oxidized silicon wafers (n-doped, resistivity of 0.008-0.02 $\Omega$cm, oxide thickness of 230-490 nm) serve as the substrate. Gate contact is made to the wafer backside. Patterned directly on the $SiO_2$ surface are chromium/gold source and drain contacts as well as additional electrodes for four-point measurements. The contacts and electrodes are defined by electron-beam lithography or photolithography using a lift-off process. A thin (5 nm thick) Cr layer is deposited first to improve adhesion, followed by 9 nm of Au which form the hole-injecting contacts to the organic crystal. For simpler two-terminal structures Au contacts are thermally evaporated and patterned on the substrate using a shadow mask.

Afterwards the substrates are cleaned in hot acetone, hot isopropanol and sulfuric peroxide, and the gate insulator and the contact surfaces are chemically treated with molecules that form self-assembled monolayers. This two-step treatment process has been improved recently to ensure a more reproducible substrate surface quality. In a controlled atmosphere, first the gate-insulator surface is chemically treated by immersing the substrates in a 3 mMol solution of octadecyltrichlorosilane (OTS) in toluene.[14] This treatment, originally developed to improve the performance of OTFTs,[15] also reduces trapping in the single crystal OFETs, as we reported previously.[10] Secondly, the substrates are immersed in a 1 to 10 mMol solution of trifluorobenzenethiol in ethanol to chemically treat the contacts for improved carrier injection/reduced contact resistance. The trifluorobenzenethiol is expected not to chemisorb on the OTS-treated $SiO_2$ surface. Preliminary contact angle measurements show this to be true for the thiols we have investigated for contact treatment. This treatment has been shown to greatly improve device performance of OTFTs and OFETs, notably in the linear region of operation where contact effects can strongly limit the performance.[10, 16] The single crystal OFETs are completed by placing a thin (<5 μm thick) OMC on top of the patterned contacts. This is done in room air under microscope illumination. A simplified top view schematic of the completed device structure is shown in Fig. 1a. The transistor channel length $L$ and width $W$ are defined by the separation and the width of the source and drain contacts, respectively.

Prior to growing the single crystals all starting materials were purified twice by temperature-gradient vacuum sublimation. The purified material was then used to grow single crystals by physical vapor transport in high-purity argon, with some additional purification resulting from the growth process itself.

Preliminary x-ray diffraction studies of our thin planar rubrene crystals reveal the crystals are orthorhombic with lattice parameters $a = 7.17$ Å, $b = 14.41$ Å, $c = 26.90$ Å, $\alpha = \beta = \gamma = 90°$, and $Z = 4$. These lattice parameters are in close agreement to previously reported values.[17, 18] However, there is some disagreement in the literature as to the space-group of single crystal rubrene. Our preliminary study does not allow us to unambiguously decide whether the crystals are space group Bbam or Aba2 as reported in references [17] and [18], respectively. The in-plane transport direction of charge in the field-accumulated carrier channel of the FET corresponds to the *ab*-plane of the crystal for the lattice parameters defined above. This is also the case for pentacene and tetracene, for which we find the same crystal structure and lattice parameters that have been reported in the literature.[19, 20]

Guarded two- and four-terminal electrical measurements were carried out in an argon glove box using a HP 4155A Semiconductor Parameter Analyzer and a HP 41501A Pulse Generator Unit. For the two-terminal measurements, the mobility $\mu$ of the field-induced carriers was evaluated using relationships developed to describe the drain current $I_D$ of single crystal silicon insulated gate FETs. In the linear region of device operation, $|V_{DS}| < |V_{GS}-V_T|$ (where $V_T$ is the threshold voltage, $V_{GS}$ is the gate-source voltage, and $V_{DS}$ is the drain-source voltage), the field-effect mobility is given by

$$\mu = \frac{L}{W} \times \frac{1}{C_i} \times \frac{1}{V_{DS}} \times \frac{\partial I_D}{\partial V_{GS}} \tag{1}$$

where $C_i$ is the gate insulator capacitance per unit area.[21]

In the saturation region of operation, where $|V_{DS}| > |V_{GS}-V_T|$, the field-effect mobility is given by[21]

$$\mu = 2 \times \frac{L}{W} \times \frac{1}{C_i} \times \left(\frac{\partial \sqrt{I_D}}{\partial V_{GS}}\right)^2 \tag{2}$$

In the four-terminal measurements (see Fig. 1), the voltage drops $V_{c1}$ and $V_{c2}$ between the source contact and the two voltage-probing electrodes were measured separately. As shown in Fig. 1b (an idealized depiction of the change in potential through the device when biased in the linear region of device operation), the potential drop between the voltage probes is assumed to be linear for $|V_{DS}| < |V_{GS}-V_T|$. The channel conductivity $\sigma$ is then given by

$$\sigma = \frac{I_D}{V_{c2} - V_{c1}} \times \frac{L'}{W} \tag{3}$$

where $L'$ is the inter-electrode spacing between the two voltage probes.

We define an "effective gate-voltage" $V_G$ between the gate contact and the channel region between the two voltage probes, where the channel conductivity is measured, as

$$V_G = V_{GS} - (V_{c1} + V_{c2})/2 \tag{4}$$

Above threshold ($V_G > V_T$) the conductivity increase was observed to be essentially linear in $V_G$ with the untrapped charge per unit area in the channel given by

$$p_{free} \cdot e = C_i \cdot (V_G - V_T) \tag{5}$$

Since $\sigma = p_{free} \cdot e \cdot \mu$, this leads to the following expression for the mobility $\mu$ of the field-induced charge carriers extracted from a four-terminal measurement:

$$\mu = \frac{\sigma}{p_{free} \cdot e} = \frac{1}{C_i} \cdot \frac{\partial \sigma}{\partial V_G} \tag{6}$$

*Results and Discussion* – We discuss first the evaluation of four-terminal measurements on a rubrene FET. Separating experimentally the contact effects from the channel effects, these measurements allow for a detailed investigation of the charge carrier mobility and of trapping in the channel area. The two-terminal output characteristics ($I_D$-$V_{DS}$) of the rubrene FET (sample A) are shown in Fig. 2a. The high quality SiO$_2$ results in a measured gate current that is several orders of magnitude smaller than the drain current. The OFET geometry is $L$ = 16 μm, $W$ = 500 μm, and the silicon dioxide is $d_{ox}$ = 300 nm thick. The two voltage probes used for the four-terminal measurements are spaced 2.75 μm from the source and drain contacts, giving an inter-electrode distance (center to center spacing) $L'$ of 10.5 μm. The current-voltage characteristics show the transition from the linear regime to the saturation regime of device operation. The curvature in the characteristics at low $V_{DS}$ indicates contact effects that should be taken into account when extracting the charge carrier mobility. From the transfer characteristics we extract a threshold voltage of –10 ± 2 V.[22]

Plotted in Fig. 2b is the voltage difference between the voltage probes ($V_{c2}$-$V_{c1}$) from the same (four-terminal) measurement as in Fig. 2a. This voltage difference varies linearly with increasing $V_{DS}$ over a limited range of $V_{DS}$ and shows almost no gate voltage dependence. Such linear dependence on $V_{DS}$ is expected if the device behaves as predicted by simple FET theory and the voltage probes in the channel are far from the contacts, i.e. not influenced by contact effects. At large $V_{DS}$ the potential difference between the voltage probes electrodes is nearly constant and independent of $V_{DS}$. This behavior is also predicted from simple FET theory for the saturation region, where one expects a nearly constant voltage drop in the channel between the source contact and the edge of the depletion region near the drain contact.

The extraction method for the mobility of the four-terminal measurement is validated by the agreement of $V_{c2}$-$V_{c1}$ with simple FET theory as shown in Fig. 2b. The part of the channel reflected in the voltage difference $V_{c2}$-$V_{c1}$ roughly corresponds to 2/3 of the channel length, given by $L'/L$. From Fig. 2b we estimate that in the linear region for $V_{DS}$ = -10 V (the applied voltage between the source and drain contacts) the voltage drop across the intrinsic channel of the FET is ~ 5.5 V. Correspondingly, the total voltage drop at the contacts is ~ 4.5 V, implying that the combined contact resistance is comparable to the channel resistance and thus significant despite the chemical treatment of the contacts.

The simplified depiction in Fig. 1b explicitly shows the voltage drop at the source and drain contacts (labeled $V_{Cont,S}$ and $V_{Cont,D}$). $V_{Cont,S}$ and $V_{Cont,D}$ are taken to be the same, although they may be different depending on the charge injection/extraction processes and interface properties. In OTFTs, asymmetry in the voltage drop at the source and drain contacts has been observed experimentally by scanning potential imaging and scanning Kelvin probe measurements.[23, 24] For our rubrene device, the contact resistances at the source and the drain contact, $R_{Cont,S}$ and $R_{Cont,D}$, have been extracted from the transfer characteristics, and are given by

$$R_{Cont,S} = \left(V_{c1} - (V_{c2} - V_{c1}) \cdot \frac{L_1}{L'}\right) \cdot I_D^{-1} \quad (7)$$

and

$$R_{Cont,D} = \left(V_D - V_{c2} - (V_{c2} - V_{c1}) \cdot \frac{L_2}{L'}\right) \cdot I_D^{-1} \quad (8)$$

where $L_1$ and $L_2$ are the distances between the source and the electrode measuring $V_{c1}$, and the drain and the electrode measuring $V_{c2}$, respectively. For the given sample their position is approximately symmetric to the middle of the channel, i.e. $L_1 \approx L_2$.

Both contact resistances show a nontrivial functional dependence on the gate voltage, decreasing with increasing $V_{GS}$. According to the above analysis, $R_{Cont,D}$ tends to be smaller than $R_{Cont,S}$. However, the finite width of the voltage probes can lead to uncertainties in the absolute values of $R_{Cont,S}$ and $R_{Cont,D}$. $R_{Cont,D}$ is found to be essentially independent of $V_{DS}$ at high gate bias; the value of $R_{Cont,S}$ for high $V_{GS}$ and different $V_{DS}$ is shown in Fig. 3. The source contact resistance slightly decreases with increasing source-drain voltage, a behavior consistent with the Schottky-like barriers that can be expected to form at the metal/organic interfaces.

Plotted in Fig. 4 are the transfer characteristics of the rubrene device. While not reported on in detail, hysteresis between the forward and reverse measurement directions (indicated in Fig. 4 with up and down arrows) is frequently observed in the current-voltage characteristics of organic semiconductor devices, and can complicate the parameter extraction. The underlying non-equilibrium processes are often attributed to trapping in deep level states. *Lang et al.* have recently reported that even high quality organic single crystals can have broad distributions of states below/above the bandgap of ~ 1 eV width.[25] How pronounced the hysteresis is in the current-voltage measurement often depends, apart from aging effects, on the measurement sweep conditions such as the voltage step or the integration time.

For the rubrene FET the difference in mobile charge density Δp between the forward and the reverse sweep direction is estimated from $\sigma = p_{free} \cdot e \cdot \mu$ and the mobility values extracted from the linear region (see below). A high measurement speed (integration time 640 μs per voltage step $\Delta V_{GS} = 0.5$ V, corresponding to a measurement speed of near 0.1 s per voltage step) was chosen to minimize trapping during the voltage sweep and to reduce the possible formation of bias-stress induced defects such as those recently reported on for pentacene by *Lang et al.*[26] For $V_{SD} = -5$ V and a total induced charge density of the forward sweep on the order of $10^{12}$ cm$^{-2}$, Δp is as high as ~ 6 %. For a slower measurement speed (integration time 20 ms, corresponding to roughly 0.5 s per voltage step), the trapped charge density reaches ~ 17 % of the total induced charge density. Trapped charge densities on the order of $10^{11}$ cm$^{-2}$ are also typical for other samples investigated.[27]

From the transfer characteristics and the measured voltage difference between the voltage-probing electrodes the channel conductivity $\sigma$ and the hole mobility $\mu$ are extracted using the relationship given in Equation 3. Because of the hysteresis in the measurement $I_D$ was averaged for the forward and reverse sweep directions.

The hole mobilities $\mu_{2T,lin}$ and $\mu_{4T,lin}$ calculated using the two- (open symbols) and four-terminal (filled symbols) methods are plotted in Fig. 4 as a function of $V_G$ and for different $V_{DS}$. The apparent difference (by a factor of 2) in the extracted "mobility" from the two- and four-terminal methods corresponds to our estimations for the voltage drops at the contacts that reduce the "effective" $V_{DS}$ and highlights the need to use a four-probe technique to calculate reliable values for $\mu$. Whereas $\mu_{4T,lin}$ is relatively $V_G$-independent, $\mu_{2T,lin}$ appears to decrease with increasing $V_G$. This is an artifact caused by the contacts, which is not accounted for in eq. 1.

However, even for the mobility $\mu_{4T,lin}$ derived from the four-probe measurement we observe a dependence on $V_{DS}$ (for $V_{DS} < -15$ V). This $V_{DS}$-dependence is not due to errors introduced into the evaluation by hysteresis effects and scatter (which are estimated to be ~ 5%); its origin is presently under investigation. A similar dependence on $V_{DS}$ was also reported by *Podzorov et al.*[9]

To further illustrate the general usefulness of the "flip-crystal" method we list in Table I the calculated field-effect mobility and the threshold voltage for 14 rubrene, tetracene and pentacene single crystal FETs (including the device described above). An oxide thickness $d_{ox}$ of 300 nm has been used for all devices except for samples G and M. Several of the samples were four-terminal structures. To minimize uncertainities due to geometry errors, we only present four-terminal samples for which the overlap of the crystal with the voltage-measuring electrodes is 10 % or less of the total channel width.

For all samples, parameters were extracted from the transfer characteristics using two- and four-terminal measurements as described. The saturation mobilities $\mu_{2T,sat}$ given in Table I are the maximum values obtained for each sample from a conservative fit (either forward or reverse voltage sweep), whereas the linear region mobilities $\mu_{4T,lin}$ and $\mu_{2T,lin}$ were again averaged for the forward and reverse sweep directions.

The threshold voltages were extracted from the $I_D^{1/2}$-$V_{GS}$ characteristics at high $V_{DS}$. For most samples, $V_T$ varied by ± 2 V upon changing $V_{DS}$; furthermore, the value of $V_T$ was influenced by hysteresis effects. Additional shifts in $V_T$ as a result of variations in the measurement speed were minor (typically 0 to 4 V). For sample J we list two different $V_T$ values (for the forward and reverse voltage sweep) since this sample showed pronounced hysteresis at high $V_{DS}$. However, at low $V_{DS}$ hysteresis was similar to the other tetracene samples.

Apparently $V_T$ depends on the material, with tetracene having large negative values and rubrene having small negative or near zero values. For pentacene the total number of samples recently prepared is too small to comment on. Since the substrates were prepared in batches, we rule out significant substrate effects as a reason for these differences and suggest the differences to be specific for the various crystals.

For the samples presented in Table I, $\mu_{2T,lin}$ can be up to a factor of six smaller than $\mu_{2T,sat}$, a fact that is commonly observed in OTFTs and attributed to contact effects.[16, 28] The reason for the large difference in $\mu_{4T,lin}$ and $\mu_{2T,sat}$ for sample B is presently unclear. The tetracene devices in general had large negative $V_T$. To limit bias stress effects we limited the magnitude of the gate bias that was applied during the measurement. The mobility therefore had to be extracted at low $V_{DS}$ and the values given in Table I should be regarded as a lower limit for the mobility in the linear region.

The $I_D^{1/2}$-$V_{GS}$ characteristics of three of the rubrene FETs, samples C, F, and G (but not the highest mobility device, sample E), showed a transition from a high mobility range at low gate bias to a lower mobility range at high gate bias (Fig 6). In Table I, we therefore indicate two different $\mu_{2T,sat}$ values for these samples corresponding to the two different regimes. The $V_T$ value in Table I is the threshold voltage of the high-mobility part of the curve, which is several volts lower than $V_T$ of the low-mobility region. The gate bias at which the crossover occurs varied from sample to sample (ca. 2-15 V above threshold). Upon changing $V_{DS}$ this crossover gate voltage remained essentially unchanged, only a smoothing of the transition could partly be observed. The origin of this gate potential-dependent (not current-dependent) effect causing the crossover is currently under investigation. However, we can rule out the transition to be an artifact caused by the measurement setup

The highest saturation field-effect mobilities measured so far are 10.7 cm$^2$/Vs for rubrene, 1.3 cm$^2$/Vs for tetracene and 1.4 cm$^2$/Vs for pentacene FETs. For pentacene, this value is nearly a factor of three higher than the best previously reported single crystal OFET results[10, 11], but still considerably lower than the best OTFT mobilities.[5] In the case of rubrene, our values are slightly higher than previously published single crystal results.[9] To the best of our knowledge, no high mobility rubrene OTFTs have been reported. For tetracene, the best OTFTs reported on to date have mobilities more than an order of magnitude lower than ours[6], whereas the best previously reported single crystal field-effect mobilities[13] are lower by a factor of three.

The transfer characteristics of the rubrene sample with the highest mobility, sample E, are plotted in Fig. 7 on a semi-logarithmic scale. The on-off ratio is as large as 10$^7$ and thus higher than the best previously reported values for rubrene single crystal FETs.[9, 29] The subthreshold swing decreases with decreasing $V_{DS}$ and reaches 0.7 V/decade at $V_{DS}$ = -5 V. This corresponds to a normalized subthreshold swing of 8 V·nF/decade·cm$^2$ ($C_i$ = 11.5 nF/cm$^2$). This value is comparable to what has been reported for the best pentacene and tetracene TFTs[30, 6] as well as pentacene single crystal FETs[11]. For tetracene and rubrene single crystal field-effect devices[12, 9] on the other hand, the smallest normalized subthreshold swings that have been published are even lower by a factor of 3 to 5 than the value we report.

Most significant is the excellent reproducibility of the results. For example, out of the last ten tetracene samples we studied, nine had a maximum saturation field-effect mobility of at least 0.35 cm$^2$/Vs, of which five had values of 0.7 cm$^2$/Vs or higher. Four of the last ten rubrene samples four showed maximum saturation mobilities of 3.4 cm$^2$/Vs or more, and the other six had mobilities greater than 1 cm$^2$/Vs. The fact that $\mu_{2T,sat}$ is higher for samples with a smaller $V_T$ is a further sign of the quality of these high-mobility devices. Whereas large sample-to-sample deviations have been reported previously for SCLC measurements on tetracene single crystals[8], resulting from defects at the metal/organic surface induced during the fabrication of the metal contacts, our FET layout avoids the problem of metal deposition on the crystal surface. The present experiments indicate that the reproducibility of the FET results and the reliability of the fabrication process is mainly due to an improved and reproducible substrate quality, namely to the modified cleaning and chemical treatment processes. FET performance can be limited by two effects, the quality of the insulator and the contact interfaces as well as the quality of the crystal itself. We have modified the device fabrication to minimize the first effect. For the organic single crystals reported on in this study, only a limited number of purification and growth steps was used, and the crystals were placed on the substrates in room air under microscope illumination. Thus the crystal surface may also have deteriorated due to photooxidation.[31] Since high trap concentrations have been reported for crystals grown from material that has been prepurified two or three times[25, 11, 12], considerable improvement in the device performance can be expected if the chemical and structural defects are reduced. Work towards a better crystal quality is currently in progress.

*Conclusion* – We have shown recent improvements in organic single crystal field-effect devices fabricated using a "flip-crystal" technique which reduces crystal handling and the likelihood of damaging the crystal surface where charge is accumulated and transported. A reliable device fabrication process with a modified chemical substrate treatment has led to a good reproducibility of the FET results. The highest hole mobilities measured so far are 10.7 cm$^2$/Vs for rubrene, 1.3 cm$^2$/Vs for tetracene, and 1.4 cm$^2$/Vs for pentacene. Four-terminal measurements allowed for the separation of contact effects from the channel resistance and thus for a reliable evaluation of the mobility. Hysteresis, which is still present in these high-mobility devices, could thus be attributed to trapping in deep level states on the order of $10^{11}$ cm$^{-2}$. Whereas the mobilities extracted from four-terminal measurements are found to be independent of $V_{GS}$, a dependence on $V_{DS}$ is observed for low $V_{DS}$. The source of this dependence of the mobility on $V_{DS}$ in the four-terminal measurement is a subject of further study. The "flip-crystal" approach is shown here to be well-suited as a general method for studying the intrinsic charge transport properties of suitably grown OMCs, and further improvement is expected from more rigorous material purification and the growth of higher quality crystals.

**Acknowledgments**

The authors thank K. Mattenberger and H.-P. Staub for assistance with crystal growth, device fabrication, and characterization, and Professor K. Ensslin and his group for the use of their laboratory facilities for substrate preparation. We also thank the group of Dr. J. Gobrecht at the Paul Scherrer Institute Villigen for providing silicon substrates with high quality thermally grown oxide, and Dr. T. Siegrist for help in early crystallography work. The authors gratefully acknowledge financial support from the Swiss National Science Foundation.

**Figure Captions**

Fig. 1  a) Schematic top view of a "flip-crystal" single crystal field-effect device.

b) Schematic voltage drop along the channel for an organic field-effect transistor biased in the linear region of device operation. The potential drop between the voltage-probing electrodes is assumed to be linear for the evaluation of the mobility from the four-terminal measurement.

Fig. 2  a) Output characteristics ($I_D$-$V_{DS}$) of a rubrene single crystal field-effect transistor (sample A; L = 16 μm, W = 500 μm, oxide thickness $d_{Ox}$ = 0.3 μm).

b) Corresponding voltage drop $V_{c2}$-$V_{c1}$ between the voltage-probing electrodes for different $V_{GS}$.

Fig. 3  Normalized contact resistance of the source contact ($R_{Cont,S}·W$) of sample A at high $V_{GS}$. The contact resistance slightly decreases with increasing source-drain voltage, which is in agreement with the Schottky-type behaviour that can be expected for a metal/organic semiconductor contact.

Fig. 4  OFET transfer characteristics ($I_D$-$V_{GS}$) for sample A. Hysteresis effects due to trapping between the forward and the reverse sweep are often observed in OFETs.

Fig. 5  Hole mobility of the rubrene single crystal device (sample A) in the linear region. Filled symbols: mobility extracted from the four-terminal measurement; open symbols: mobility extracted from the two-terminal measurement.

Fig. 6  $I_D^{1/2}$-$V_{GS}$ characteristics for a rubrene single crystal FET (sample C) showing a transition from a high-mobility region at low gate bias to a lower-mobility region at larger $V_{GS}$.

Fig. 7  Transfer characteristics (forward and reverse voltage sweep) for the rubrene single crystal FET with the highest mobility (sample E, L = 16.5 μm, W = 265 μm, $d_{Ox}$ = 0.3 μm).

Table I: Summary of single crystal FET characteristics for rubrene, tetracene and pentacene samples.

| Sample | Single crystal | W/L (μm/μm) | $\mu_{4T,lin}$ (cm$^2$/Vs) | $\mu_{2T,lin}$ (cm$^2$/Vs) | $\mu_{2T,sat}$ (cm$^2$/Vs) | $V_T$ (V) |
|---|---|---|---|---|---|---|
| A | rubrene | 500/16 | 1.4 | 0.7 | 1.8 | -10 |
| B | rubrene | 16.5/503 | 0.9 | 0.8 | 3.4 | -2 |
| C | rubrene | 505/16.5 | 1.5 | 1.3 | 5.5 (1.6) | 1 |
| D | rubrene | 81/16 | 1.8 | 0.7 | 1.5 | -12 |
| E | rubrene | 265/16.5 | | 1.9 | 10.7 | 0 |
| F | rubrene | 500/16 | | 1.0 | 3.9 (1.5) | -4 |
| G | rubrene | 505/16.5 | | 0.9 | 3.6 (1.9) | -5 |
| H | pentacene | 83/75 | | 0.8 | 1.4 | 12 |
| I | pentacene | 208/16.5 | | 0.6 | 0.9 | 11 |
| J | tetracene | 105/16.5 | | 0.1 | 1.3 | -7 (forward) -21 (reverse) |
| K | tetracene | 268/16.5 | | 0.04 | 0.8 | -34 |
| L | tetracene | 500/16 | | 0.1 | 0.7 | -22 |
| M | tetracene | 139/16 | | 0.3 | 0.8 | -39 |
| N | tetracene | 395/17 | | 0.3 | 0.8 | -14 |

a)

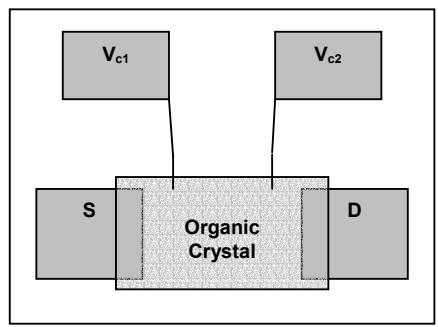

b)

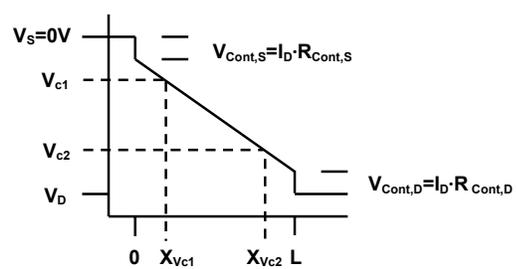

*Fig. 1 of 7, C. Goldmann et al.*

a)

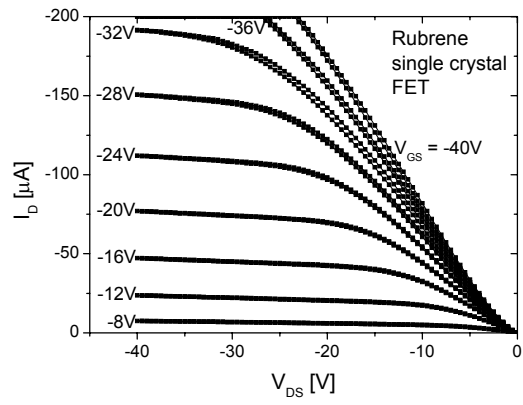

b)

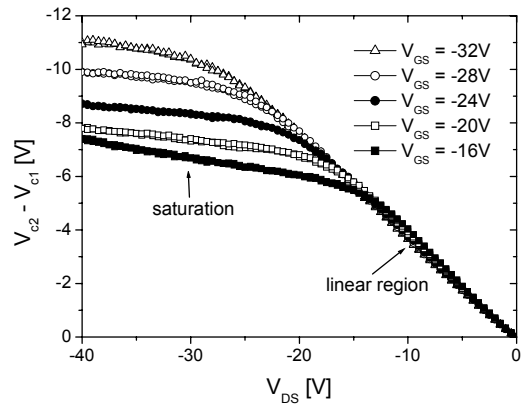

*Fig. 2 of 7, C. Goldmann et al.*

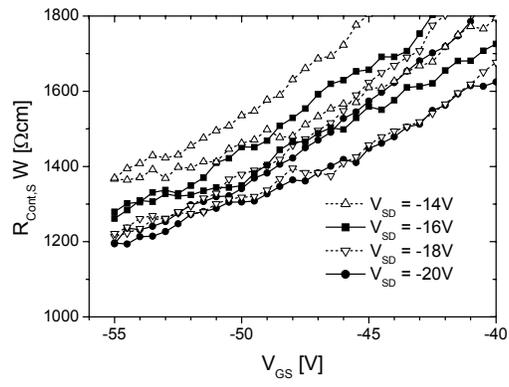

*Fig. 3 of 7, C. Goldmann et al.*

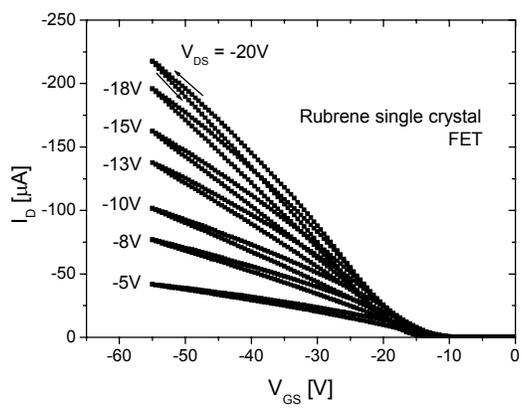

*Fig. 4 of 7, C. Goldmann et al.*

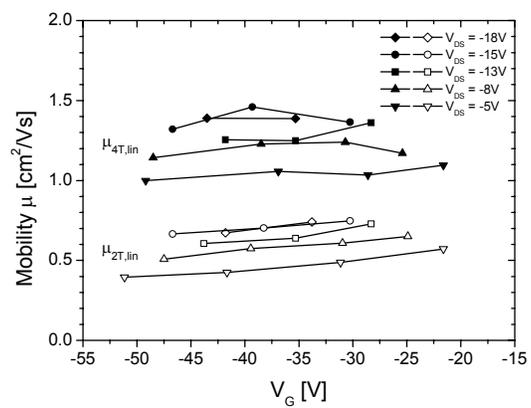

*Fig. 5 of 7, C. Goldmann et al.*

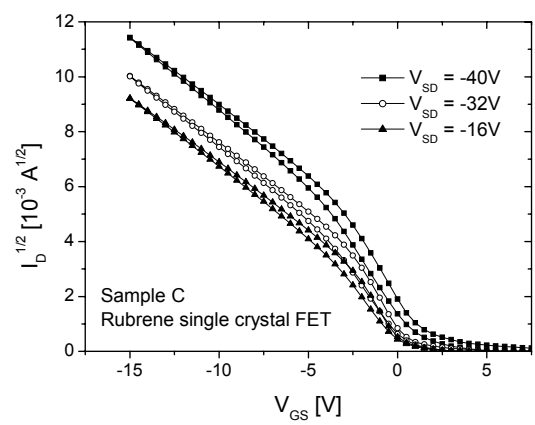

*Fig. 6 of 7, C. Goldmann et al.*

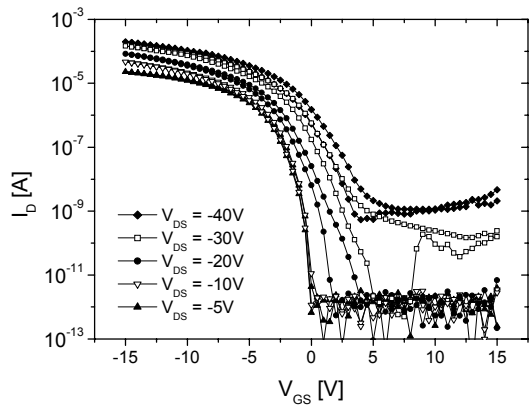

*Fig. 7 of 7, C. Goldmann et al.*